\documentclass[a4paper,aps,prd,twocolumn,tightenlines,preprintnumbers,nofootinbib,showkeys,superscriptaddress]{revtex4-1}
\pdfoutput=1

 \usepackage{XCharter}
 \usepackage[T1]{fontenc}
 \usepackage{mathptmx}

\usepackage{fullpage}
\usepackage{amsfonts}
\usepackage{epic}
\usepackage{eepic}
\usepackage{epsfig}
\usepackage{latexsym}
\usepackage{graphicx,slashed,xcolor,multirow,amsmath,amssymb}
\usepackage{tikz}
\usepackage{tikz-feynman}
\usepackage{caption}
\usepackage{textcomp}
\usepackage{subcaption}
\usepackage{adjustbox}

\usepackage{float}
\usepackage{multirow}
\usepackage{hyperref}
\usepackage{enumitem}

\usepackage{natbib}
\usepackage{relsize}

\usepackage[left=1.75cm, right=1.75cm, top=2.1cm, bottom=2.5cm, headsep=1cm]{geometry}
\begin{document}
\title{Exploring $\widetilde{R}_2$ Leptoquarks and Majorana Neutrinos via same-sign dimuons at the HL-LHC.}

\author{Subham Saha}\email{subham.saha@iopb.res.in}
\affiliation{Institute of Physics, Sachivalaya Marg, Bhubaneswar, Odisha 751005, India}
\affiliation{Homi Bhabha National Institute, BARC Training School Complex, Anushakti Nagar, Mumbai
400094, India}

\author{Arvind Bhaskar}\email{arvind.bhaskar@iopb.res.in}
\affiliation{Institute of Physics, Sachivalaya Marg, Bhubaneswar, Odisha 751005, India}
\affiliation{Homi Bhabha National Institute, BARC Training School Complex, Anushakti Nagar, Mumbai
400094, India}

\author{Manimala Mitra}\email{manimala@iopb.res.in}
\affiliation{Institute of Physics, Sachivalaya Marg, Bhubaneswar, Odisha 751005, India}
\affiliation{Homi Bhabha National Institute, BARC Training School Complex, Anushakti Nagar, Mumbai
400094, India}

\begin{abstract}
\noindent
We study the phenomenology of the scalar leptoquark (sLQ) $\widetilde{R}_2$ coupled to right-handed neutrinos (RHNs) at the High-Luminosity Large Hadron Collider (HL-LHC), focusing on signatures beyond those targeted in conventional sLQ searches. In the regime where the sLQ is heavier than the RHN and the Yukawa couplings are $\mathcal{O}(1)$, the decay $\widetilde{R}_2 \to N j$ can dominate, leading to distinctive same-sign dimuon and multijet final states. This lepton-number-violating signature is weakly constrained by existing searches, benefits from low Standard Model backgrounds, and provides a direct probe of the Majorana nature of the RHN. We perform a comprehensive analysis by combining sLQ pair and single production at $\sqrt{s}=14~\text{TeV}$, and evaluate the HL-LHC sensitivity over a wide range of masses and Yukawa couplings. We find that pair production dominates at the TeV scale, while single production becomes increasingly important at higher masses, significantly extending the accessible parameter space beyond current direct and indirect limits. Our results demonstrate the strong complementarity between production modes and highlight the potential of the HL-LHC to probe sLQ scenarios with Majorana RHNs.

\end{abstract}
\preprint{IOP/BBSR/2026-04}
\maketitle

\section{Introduction}
\noindent
Leptoquarks (LQs) constitute a well-motivated class of hypothetical particles that arise in a wide range of theoretical frameworks aiming at a unified description of quarks and leptons~\cite{Buchmuller:1986zs,Blumlein:1994qd,Blumlein:1996qp,Dorsner:2016wpm}. As color-triplet bosons carrying both baryon and lepton quantum numbers, they provide a natural link between the two fermionic sectors of the Standard Model (SM). LQs appear generically in various beyond the Standard Model (BSM) scenarios, including Pati-Salam models~\cite{Pati:1974yy}, $\mathrm{SU}(5)$ grand unified theories~\cite{Georgi:1974sy}, models with quark-lepton compositeness~\cite{Schrempp:1984nj}, $R$-parity violating supersymmetric models~\cite{Barbier:2004ez}, and the coloured Zee-Babu model~\cite{Kohda:2012sr}. Beyond their theoretical appeal, LQs have received renewed phenomenological attention due to their potential to address persistent experimental anomalies, notably in semileptonic $B$-meson decays~\cite{Queiroz:2014pra,HeavyFlavorAveragingGroupHFLAV:2024ctg} and in the anomalous magnetic moment of the muon~\cite{Queiroz:2014zfa,Aoyama:2020ynm}.

The Large Hadron Collider (LHC) has played a central role in the experimental search for LQs. The ATLAS and CMS collaborations have set stringent exclusion limits on their masses, extending up to the TeV scale for a variety of LQ representations. These searches are primarily optimized for final states with high-transverse-momentum leptons and jets, corresponding to the conventional LQ decay into a charged lepton and a quark~\cite{ATLAS:2020dsk}. Complementary indirect constraints on LQ couplings to SM fermions arise from high-$p_T$ dilepton and monolepton plus missing transverse momentum searches~\cite{Bessaa:2014jya,Mandal:2018kau,ATLAS:2020yat,Babu:2020hun,Bhaskar:2021pml,Angelescu:2021lln,CMS:2023qdw}. Nevertheless, several phenomenologically well-motivated decay modes remain only weakly constrained. In particular, scenarios in which LQs couple to right-handed neutrinos (RHNs) are of special interest~\cite{Mandal:2018qpg,Padhan:2019dcp,Cottin:2021tfo,Bhaskar:2023xkm,Varzielas:2023qlb,Desai:2023jxh,Duraikandan:2024kcy}. RHNs provide a minimal and well-established extension of the SM that can naturally account for the smallness of neutrino masses via the seesaw mechanism~\cite{Minkowski:1977sc, Gell-Mann:1979vob, Mohapatra:1979ia, Yanagida:1979as, Yanagida:1980xy}. 

\noindent
If the sLQ is heavier than the associated RHN, the decay of the sLQ into an RHN and a jet can dominate. This scenario significantly alters the expected collider phenomenology, by modifying existing bounds derived from conventional LQ searches and opening up a largely unexplored region of parameter space. The subsequent decay of the RHN can give rise to rich multi-lepton and multi-jet final states. While our earlier work explored the discovery prospects of such sLQs at a future high-energy muon collider~\cite{Saha:2025npi}, the High-Luminosity LHC (HL-LHC)~\cite{Apollinari:2015wtw} provides an immediate and complementary opportunity to probe these scenarios, owing to its large projected integrated luminosity.

In this work, we focus on a particularly clean and striking signature: the production of sLQs followed by decays into RHNs. If the RHN is Majorana in nature, it can induce same-sign dimuon final states accompanied by multiple jets, $\mu^\pm \mu^\pm + \text{jets}$. This final state is characterized by exceptionally low SM backgrounds, making it a powerful probe of lepton-number-violating new physics. The observation of a statistically significant excess in same-sign dimuon events would therefore constitute a smoking-gun signature of both sLQ interactions and the Majorana nature of the RHN. We perform a dedicated HL-LHC study of this channel, developing an optimized search strategy and evaluating the discovery potential for TeV-scale sLQs coupled to RHNs.

\noindent
The rest of this paper is organized as follows. In Section~\ref{sec:2}, we introduce the sLQ model under consideration and define the relevant interactions. Section~\ref{sec:3} discusses the decay modes of the sLQ and RHN. In Section~\ref{sec:4}, we describe the sLQ production mechanisms at the HL-LHC relevant to the same-sign dimuon and multijet final state. The signal topology, kinematic distribution, SM backgrounds, and event selection strategy are presented in Section~\ref{sec:5}. The statistical framework used to evaluate the signal significance is outlined in Section~\ref{sec:6}. Our results are presented in Section~\ref{sec:7}, followed by our conclusions in Section~\ref{sec:8}.

\section{Model}
\label{sec:2} 
\noindent
To facilitate our analysis, we extend the SM by introducing two types of BSM particles: sLQ doublet, $\widetilde{R}_2$ with quantum numbers $({\bf 3},{\bf 2},1/6)$ and three generations of RHN, which are singlet $({\bf 1},{\bf 1},0)$ under the SM. The isospin components of the $\widetilde{R}_2$ doublet can be written as $\widetilde{R}_2 = (\widetilde{R}_2^{2/3}, \widetilde{R}_2^{-1/3})$. The most general, renormalizable Lagrangian describing the interactions of the $\widetilde{R}_2$ sLQ with SM fermions and the RHN, following the notation in~\cite{Dorsner:2016wpm,Buchmuller:1986zs,Padhan:2019dcp}, is given as: 
\begin{align}
\mathcal{L}_{\rm LQ} = - Y_{ij}\bar{d}_{R}^{i}\widetilde{R}_{2}^{a}\epsilon^{ab}L_{L}^{j,b}+
Z_{ij}\bar{Q}_{L}^{i,a}\widetilde{R}_{2}^{a}N_{R}^{j}+\text{H.c.} \, ,
\label{eq:eq1}
\end{align}
Here $i, j = 1, 2, 3$ are the generation indices, $a, b = 1, 2$ are $SU(2)_L$ indices and $\epsilon^{ab}$ is the $SU(2)$ antisymmetric tensor. The matrices $Y_{ij}$ and $Z_{ij}$ contain the Yukawa couplings. When expanded, this Lagrangian yields the interaction terms for the component fields: 
\begin{align}
&\mathcal{L}_{\rm LQ} =-Y_{ij}\bar{d}_{R}^{i}e_{L}^{j}\widetilde{R}_{2}^{2/3}+(YU_{\text{PMNS}})_{ij}\bar{d}_{R}^{i}\nu_{L}^{j}
\widetilde{R}_{2}^{-1/3} \newline\\
&+ (V_{\text{CKM}}Z)_{ij}\bar{u}_{L}^{i}N_R^{j}\widetilde{R}_{2}^{2/3}+Z_{ij}\bar{d}_{L}^{i}
N_{R}^{j}\widetilde{R}_{2}^{-1/3}+\text{H.c.}
\label{eq:eq2}
\end{align}
In the above equation, $U_{\text{PMNS}}$ and $V_{\text{CKM}}$ are the Pontecorvo-Maki-Nakagawa-Sakata (PMNS) and Cabibbo-Kobayashi-Maskawa (CKM) mixing matrices, respectively. For simplicity, we consider a scenario with minimal flavor assumptions. We consider Yukawa couplings $Y_{12}$ that quantifies the coupling strength of a sLQ, first generation quark and a second generation lepton and $Z_{11}$ that quantifies the coupling strength of a sLQ, first generation quark and a first generation RHN. The first generation RHN, assumed to be a Majorana fermion is denoted as $N$. In addition to its Yukawa interaction with the sLQ, we consider that $N$  mixes with the SM muon neutrino $\nu_{\mu}$, through an active-sterile mixing angle, $V_{\mu N}$. This mixing induces interactions between the RHN and the SM electroweak bosons ($W,Z, H$), governed by the magnitude of $V_{\mu N}$. The relevant interaction terms are:
\begin{align}
   &\mathcal{L}_{\mu W N} = \frac{g}{\sqrt{2}} W^-_{\mu} \bar{\mu} \gamma^{\mu} P_L V_{\mu N} N + {\rm H.c.},
\label{eq:WlN} \\
   &\mathcal{L}_{\nu_{\mu} Z N} = \frac{g}{2\cos\theta_w} Z_{\mu} \bar{\nu}_{\mu} \gamma^{\mu} P_L V_{\mu N} N + {\rm H.c.},
\label{eq:ZnuN} \\
   &\mathcal{L}_{\nu_{\mu} H N} = \frac{M_N}{v} H \bar{\nu}_{\mu} P_R V_{\mu N} N + {\rm H.c.}
\label{eq:HnuN}
\end{align}
Here, $M_N$ is the mass of the RHN. In this work we denote the mass of both components of $\widetilde{R}_2$ as $M_{\widetilde{R}_2^{-1/3}}=M_{\widetilde{R}_2^{2/3}}=M_{\widetilde{R}_2}$

\section{Decay modes}
\label{sec:3} 
\noindent
In this section, we briefly explain the plausible decay modes of sLQ and RHN depending on the couplings mentioned above. A detailed explanation of these decay modes and their expressions is given in~\cite{Saha:2025npi}.
\begin{enumerate}
    \item \textbf{Leptoquark Decays}: The branching ratios of $\widetilde{R}^{2/3}_2$ into the $uN$ and $d\mu$ final states as function of the Yukawa couplings have been studied in detail in~\cite{Saha:2025npi} for the different benchmark values of $M_N$. In the regime $M_{\widetilde{R}^{2/3}_2} > M_N$ and a large $Z_{11}$,  the two-body decay $\widetilde{R}^{2/3}_2 \to u N$ competes with the mode $\widetilde{R}^{2/3}_2 \to \mu d$. Owing to the large mass hierarchy between the sLQ and the SM fermions, the branching ratios are primarily governed by the relative sizes of the Yukawa couplings: the charged-lepton channel increases with $Y_{12}$ and decreases with $Z_{11}$, while the RHN channel exhibits the complementary behavior.

    \item \textbf{Right-Handed Neutrino Decays}: The RHN can decay via two qualitatively different mechanisms: through its mixing with SM neutrino, giving rise to three-body decays mediated by off-shell $W$, $Z$, and Higgs bosons for lighter $N$, and through off-shell sLQ exchange. For heavier $N$, the former decay mode will get replaced by on-shell two-body decays, such as $N \to l W, \nu_{\mu} Z, \nu_{\mu} H$. The relative importance of these channels depends crucially on the active--sterile mixing $V_{\mu N}$, the respective Yukawa couplings, and the sLQ mass. A detailed discussion of the corresponding branching-ratio patterns for representative benchmark choices (including $M_N = 50$~GeV) can be found in~\cite{Saha:2025npi}. In general, for large mixing the electroweak-mediated channels dominate and the branching ratios are largely insensitive to the LQ mass, whereas for very small mixing the LQ-mediated modes become dominant, yielding sizable hadronic final states. For intermediate mixing values, both contributions are relevant, leading to a smooth transition in the decay pattern with increasing LQ mass.

\end{enumerate}


\section{LQ Production at the HL-LHC}
\label{sec:4}
\begin{figure*}
	\centering
	
	\begin{subfigure}[b]{0.30\textwidth}
		\centering
		\begin{tikzpicture}
			\begin{feynman}
				\vertex (a) at (0,0);
				\vertex (b) at (-1.5, 1.5);
				\vertex (c) at (-1.5, -1.5);
				\vertex (f) at (2.0, 0);
				\vertex (d) at (3.5, 1.5);
				\vertex (e) at (3.5, -1.5);
				
				\diagram*  {
					(a) -- [edge label=$p$, fermion] (b),
					(c) -- [edge label=$p$, fermion] (a),
					(a) -- [edge label=$Z / \gamma$, boson] (f),
					(d) -- [edge label=$\widetilde{R}^{2/3}_2/ \widetilde{R}^{-1/3}_2$, scalar] (f),
					(f) -- [edge label=$\widetilde{R}^{-2/3}_2 / \widetilde{R}^{1/3}_2$, scalar] (e),
				};
			\end{feynman}
		\end{tikzpicture}
		\caption{}
		\label{fig:R2pairQED}
	\end{subfigure}
\hfill
\begin{subfigure}[b]{0.30\textwidth}
	\centering
	\begin{tikzpicture}
		\begin{feynman}
			\vertex (a) at (0,0);
			\vertex (b) at (-1.5, 1.5);
			\vertex (c) at (-1.5, -1.5);
			\vertex (f) at (2.0, 0);
			\vertex (d) at (3.5, 1.5);
			\vertex (e) at (3.5, -1.5);
			
			\diagram*  {
				(a) -- [edge label=$p$, fermion] (b),
				(c) -- [edge label=$p$, fermion] (a),
				(a) -- [edge label=$W^{\pm}$, boson] (f),
				(d) -- [edge label=$\widetilde{R}^{2/3}_2/ \widetilde{R}^{-2/3}_2$, scalar] (f),
				(f) -- [edge label=$\widetilde{R}^{1/3}_2 / \widetilde{R}^{-1/3}_2$, scalar] (e),
			};
		\end{feynman}
	\end{tikzpicture}
	\caption{}
	\label{fig:R2pairQED2} 
\end{subfigure}
\hfill
\begin{subfigure}[b]{0.30\textwidth}
	\centering
	\begin{tikzpicture}
		\begin{feynman}
			\vertex (a) at (0,0);
			\vertex (b) at (-1.5, 1.5);
			\vertex (c) at (-1.5, -1.5);
			\vertex (f) at (2.0, 0);
			\vertex (d) at (3.5, 1.5);
			\vertex (e) at (3.5, -1.5);
			
			\diagram*  {
				(a) -- [edge label=$p$, fermion] (b),
				(c) -- [edge label=$p$, fermion] (a),
				(a) -- [edge label=$g$, gluon] (f),
				(d) -- [edge label=$\widetilde{R}^{2/3}_2/ \widetilde{R}^{-1/3}_2$, scalar] (f),
				(f) -- [edge label=$\widetilde{R}^{-2/3}_2 / \widetilde{R}^{1/3}_2$, scalar] (e),
			};
		\end{feynman}
	\end{tikzpicture}
	\caption{}
	\label{fig:R2pairQCD}
\end{subfigure}
\begin{subfigure}[b]{0.30\textwidth}
	\centering
	\begin{tikzpicture}
		\begin{feynman}
			\vertex (a) at (0,0);
			\vertex (b) at (-1.0, 0.8);
			\vertex (c) at (1.0, 0.8);
			\vertex (f) at (0,-1.5);
			\vertex (d) at (1.0, -2.0);
			\vertex (e) at (-1.0, -2.0);
			
			\diagram* {
				(a) -- [edge label=$p$, fermion] (b),
				(c) -- [edge label=$\widetilde{R}^{+2/3}_2/\widetilde{R}^{-1/3}_2$, scalar] (a),
				(f) -- [edge label=$N/\mu/\nu_{\mu}$, fermion] (a),
				(f) -- [edge label=$\widetilde{R}^{-2/3}_2/\widetilde{R}^{+1/3}_2$, scalar] (d),
				(e) -- [edge label=$p$, fermion] (f),
			};
		\end{feynman}
	\end{tikzpicture}
	\caption{}
	\label{fig:R2pairNP}    
\end{subfigure}
\hfill
 \begin{subfigure}[b]{0.30\textwidth}
	\centering
	\begin{tikzpicture}
		\begin{feynman}
			\vertex (a) at (0,0);
			\vertex (b) at (-1.0, 0.8);
			\vertex (c) at (1.0, 0.8);
			\vertex (f) at (0,-1.5);
			\vertex (d) at (1.0, -2.0);
			\vertex (e) at (-1.0, -2.0);
			
			\diagram* {
				(a) -- [edge label=$p$, fermion] (b),
				(c) -- [edge label=$\widetilde{R}^{\pm2/3}_2/\widetilde{R}^{\mp1/3}_2$, scalar] (a),
				(f) -- [edge label=$N$, draw] (a),
				(f) -- [edge label=$\widetilde{R}^{\pm2/3}_2/\widetilde{R}^{\mp1/3}_2$, scalar] (d),
				(e) -- [edge label=$p$, fermion] (f),
			};
		\end{feynman}
	\end{tikzpicture}
	\caption{}
	\label{fig:R2pairNP2}    
\end{subfigure}
\hfill
\begin{subfigure}[b]{0.30\textwidth}
	\centering
	\begin{tikzpicture}[baseline={(current bounding box.center)}, scale=0.8]
		\begin{feynman}
			\vertex (a) at (0,0);
			\vertex (b) at (-1.2, 1.2);
			\vertex (c) at (-1.2, -1.2);
			\vertex (f) at (1.5, 0);
			\vertex (d) at (2.8, 1.2);
			\vertex (e) at (2.8, -1.2);
			\vertex (m) at (4.5, -2.2);
			\vertex (n) at (4.2, 0.0);
			
			\diagram*  {
				(a) -- [edge label=$p$, fermion] (b),
				(c) -- [edge label=$g$, gluon] (a),
				(a) -- [edge label=$u/d$, fermion] (f),
				(d) -- [edge label=$g$, gluon] (f),
				(f) -- [edge label=$u/d$, fermion] (e),
				(e) -- [edge label=${\widetilde{R}}^{2/3}_2/{\widetilde{R}}^{-1/3}_2$, scalar] (m),
				(e) -- [edge label=$N$, draw] (n),
			};
		\end{feynman}
	\end{tikzpicture}
	\caption{}
	\label{fig:R2single1}
\end{subfigure}
\begin{subfigure}[b]{0.30\textwidth}
	\centering
	\begin{tikzpicture}
		\begin{feynman}
			\vertex (a) at (0,0);
			\vertex (b) at (-1.0, 0.8);
			\vertex (c) at (1.0, 0.8);
			\vertex (f) at (0,-1.5);
			\vertex (d) at (1.0, -2.0);
			\vertex (e) at (-1.0, -2.0);
			\vertex (m) at (2.5, -3.0);
			\vertex (n) at (2.5, -1.2);

			\diagram* {
				(a) -- [edge label=$g$, gluon] (b),
				(c) -- [edge label=$g$, gluon] (a),
				(f) -- [edge label=$g$, gluon] (a),
				(f) -- [edge label=$u/d$, fermion] (d),
				(d) -- [edge label=${\widetilde{R}}^{2/3}_2/{\widetilde{R}}^{-1/3}_2$, scalar] (m),
				(d) -- [edge label=$N$, draw] (n),
				(e) -- [edge label=$p$, fermion] (f),
			};
		\end{feynman}
	\end{tikzpicture}
	\caption{}
	\label{fig:R2single3}
\end{subfigure}
\hfill
\begin{subfigure}[b]{0.30\textwidth}
	\centering
	\begin{tikzpicture}[baseline={(current bounding box.center)}, scale=0.8]
		\begin{feynman}
			\vertex (a) at (0,0);
			\vertex (b) at (-1.2, 1.2);
			\vertex (c) at (-1.2, -1.2);
			\vertex (f) at (1.5, 0);
			\vertex (d) at (2.8, 1.2);
			\vertex (e) at (2.8, -1.2);
			\vertex (m) at (4.5, -2.2);
			\vertex (n) at (4.2, 0.0);
			
			\diagram*  {
				(a) -- [edge label=$p$, fermion] (b),
				(c) -- [edge label=$g$, gluon] (a),
				(a) -- [edge label=$d$, fermion] (f),
				(d) -- [edge label=$g$, gluon] (f),
				(f) -- [edge label=$d$, fermion] (e),
				(e) -- [edge label=${\widetilde{R}}^{2/3}_2$, scalar] (m),
				(e) -- [edge label=$\mu^{-}$, fermion] (n),
			};
		\end{feynman}
	\end{tikzpicture}
	\caption{}
	\label{fig:R2single4}
\end{subfigure}
\hfill
\begin{subfigure}[b]{0.30\textwidth}
	\centering
	\begin{tikzpicture}
		\begin{feynman}
			\vertex (a) at (0,0);
			\vertex (b) at (-1.0, 0.8);
			\vertex (c) at (1.0, 0.8);
			\vertex (f) at (0,-1.5);
			\vertex (d) at (1.0, -2.0);
			\vertex (e) at (-1.0, -2.0);
			\vertex (m) at (2.5, -3.0);
			\vertex (n) at (2.5, -1.2);

			\diagram* {
				(a) -- [edge label=$g$, gluon] (b),
				(c) -- [edge label=$g$, gluon] (a),
				(f) -- [edge label=$g$, gluon] (a),
				(f) -- [edge label=$d$, fermion] (d),
				(d) -- [edge label=${\widetilde{R}}^{2/3}_2$, scalar] (m),
				(d) -- [edge label=$\mu^{-}$, fermion] (n),
				(e) -- [edge label=$p$, fermion] (f),
			};
		\end{feynman}
	\end{tikzpicture}
	\caption{}
	\label{fig:R2single6}
\end{subfigure}
\captionsetup{justification=raggedright,singlelinecheck=false}     
\caption{Representative Feynman diagrams illustrating the pair and single production of $\widetilde{R}_2$. }
\label{Fig:LQ_production}
\end{figure*}
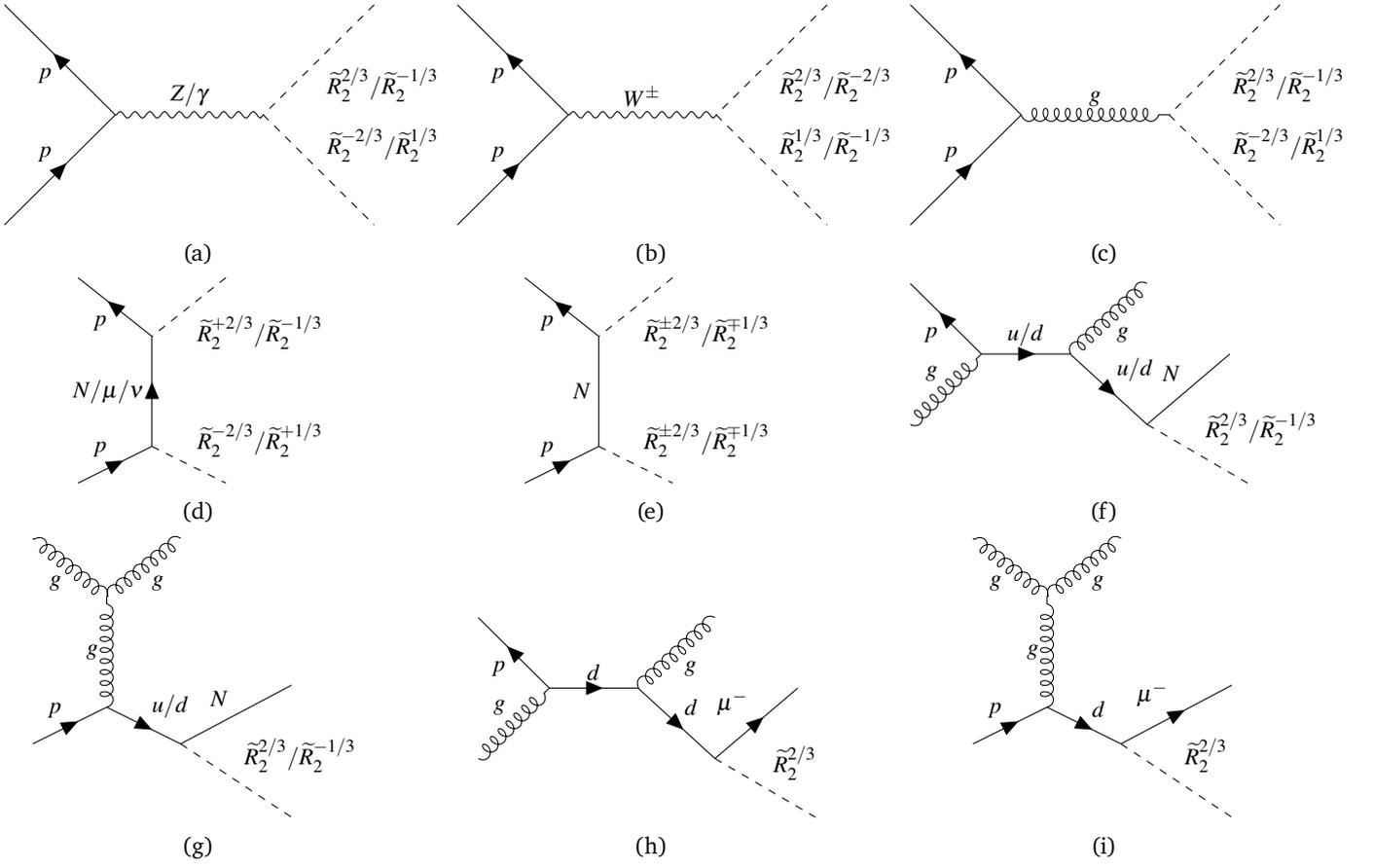

\noindent
In this section, we describe the various $\widetilde{R}_2$ production channels at the HL-LHC that yield same sign dimuon and multi-jet signatures. To realize this final state, we consider scenarios where at least one sLQ decays into a jet and a RHN. Crucially, the Majorana nature of the RHN allows it to decay into $\mu^{\pm}+$ jets; this inherent lepton number violation provides a natural mechanism for generating same sign dilepton events. Furthermore, the Majorana character of the RHN, coupled with the charge-asymmetric initial state of a proton collider, facilitates additional production modes that were not present in our previous study~\cite{Saha:2025npi}. Representative parton-level Feynman diagrams for both single and pair production are illustrated in Fig.~\ref{Fig:LQ_production}. Below, we provide an exhaustive list of all relevant pair and single production channels, together with their coupling dependence, 

\begin{enumerate}
    \item \textbf{Pair Production}: 
    \begin{itemize}
        \item [--] Similar to our previous work~\cite{Saha:2025npi}, we consider the pair production process $pp \rightarrow \widetilde{R}_2 \widetilde{R}_2^{*}  $ (where, $\widetilde{R}_2 \in \{\widetilde{R}^{2/3}_2,\widetilde{R}^{-1/3}_2 \} $), mediated by a $s$-channel $\gamma/Z$ boson or gluon. The representative Feynman diagrams for these channels are shown in Figs.~\ref{fig:R2pairQED} and \ref{fig:R2pairQCD}. In a hadron collider, we also have the mixed pair production channel $pp \rightarrow \widetilde{R}^{\pm2/3}_2 \widetilde{R}^{\pm1/3}_2$ mediated by a $s$-channel $W$ boson, as illustrated in Fig.~\ref{fig:R2pairQED2}.
        
        \item [--] In addition to the SM mediated pair production channels, pair production of $\widetilde{R}_2$ mediated by the new physics couplings ($Y_{12}$ and $Z_{11}$) is also possible. The process $pp \rightarrow \widetilde{R}^{2/3}_2 \widetilde{R}^{-2/3}_2~(\widetilde{R}^{1/3}_2 \widetilde{R}^{-1/3}_2)$ can occur via  $t$-channel [$\mu$, $N$] ([$\nu_{\mu}, N$]) (See Fig.~\ref{fig:R2pairNP}). The cross section contribution involving $t$-channel SM fermions scale as $Y^4_{12}$,  whereas the contribution involving $t$-channel $N$ scales as $Z^4_{11}$.
    
        \item[--] There is also interference between the relevant SM-mediated and new-physics–mediated diagrams. For example, interference arises between the SM mediated pair production of $\widetilde{R}^{2/3}_2 \widetilde{R}^{-2/3}_2$ and $\widetilde{R}^{-1/3}_2 \widetilde{R}^{1/3}_2$ and the corresponding processes mediated by a $t$-channel $\nu_{\mu},\mu,N$.
        The cross section contribution from these interference terms scales as $Y^2_{12}/Z^2_{11}$.
        
        \item [--] If the pair production of $\widetilde{R}_2$ proceeds only via a $t$-channel Majorana RHN, a rich set of additional pair production channels becomes accessible, all of which can lead to the same sign dimuon and multijet final state (See Fig.~\ref{fig:R2pairNP2}). These channels do not interfere with the SM-mediated processes, and their cross sections scale as $Z^4_{11}$. The relevant channels are $\widetilde{R}^{\pm 2/3}_2 \widetilde{R}^{\pm 2/3}_2, \widetilde{R}^{\pm 1/3}_2 \widetilde{R}^{\pm 1/3}_2, \widetilde{R}^{\pm 2/3}_2 \widetilde{R}^{\mp 1/3}_2$, and $\widetilde{R}^{\pm 2/3}_2 \widetilde{R}^{\pm 1/3}_2$.

        \item [--] Following pair production, sLQs are decayed according to two distinct topologies to reach the targeted final states: Symmetric mode, where both decay into a RHN and a jet, and an asymmetric mode, where one yields an RHN and a jet while the other produces a muon and a jet. These production channels are detailed in Eqs.~\ref{eq:symmetricpair} and \ref{eq:asymmetricpair}, respectively.
\end{itemize}
\item \textbf{Single Production}: In the single production mode, a $\widetilde{R}_2$ is produced alongside a quark and a SM lepton or RHN. To obtain a same sign dimuon and multijet final state in single production, at least one of the produced sLQs must decay into an RHN. We present the symmetric and asymmetric single production modes in Eqs.~\ref{eq:symmetricsingle} and \ref{eq:asymmetricsingle}, respectively. The representative Feynman diagrams for these processes are shown in Figs.~\ref{fig:R2single1}, \ref{fig:R2single3}, \ref{fig:R2single4}, and \ref{fig:R2single6}.
\end{enumerate}

\begin{center}
\textbf{Symmetric mode: Pair production}
\end{center}

\begin{widetext}
\begin{align}
\label{eq:symmetricpair}
pp\to\left\{\begin{array}{lclcl}
  \widetilde{R}^{+2/3}_2 \widetilde{R}^{-2/3}_2, \widetilde{R}^{-2/3}_2 \widetilde{R}^{-2/3}_2 \\
  \widetilde{R}^{+2/3}_2 \widetilde{R}^{+2/3}_2, \widetilde{R}^{+1/3}_2 \widetilde{R}^{-1/3}_2 \\
  \widetilde{R}^{+1/3}_2 \widetilde{R}^{+1/3}_2, \widetilde{R}^{-1/3}_2 \widetilde{R}^{-1/3}_2 \\
  \widetilde{R}^{+2/3}_2 \widetilde{R}^{-1/3}_2, \widetilde{R}^{+2/3}_2 \widetilde{R}^{+1/3}_2 \\
  \widetilde{R}^{-2/3}_2 \widetilde{R}^{-1/3}_2, \widetilde{R}^{-2/3}_2 \widetilde{R}^{+1/3}_2 \\
\end{array}\right\} \rightarrow (j N) \, (j N) \rightarrow j(\mu j j)\ j(\mu jj) \equiv \mu^{\pm} \mu^{\pm} + N_{jet} \ge 6
\end{align}
\end{widetext}
Here, $j$ could either be a light quark or an anti quark or a gluon depending on the charge of $\widetilde{R}^{2/3}_2$ or $\widetilde{R}^{-1/3}_2$.

\begin{center}
\textbf{Symmetric mode: Single Production}
\end{center}

\begin{widetext}
\begin{equation}
\label{eq:symmetricsingle}
pp\to\left\{\begin{array}{lclcl}
  \widetilde{R}^{+2/3}_2 j N,\ \widetilde{R}^{-2/3}_2 j N & \rightarrow & (j N)\ j N, (j N)\ j N  &\rightarrow& j(\mu j j)\  j(\mu jj) \equiv \mu^{\pm} \mu^{\pm} + N_{jet} \ge 6 \\
  \widetilde{R}^{+1/3}_2 j N,\ \widetilde{R}^{-1/3}_2 j N  & \rightarrow & (j N)\ j N,\ (j N)\ j N   &\rightarrow& j(\mu j j)\  j(\mu jj) \equiv \mu^{\pm} \mu^{\pm} + N_{jet} \ge 6 \\
\end{array}\right\}
\end{equation}    
\end{widetext}

\begin{center}
\textbf{Asymmetric mode: Pair production}
\end{center}

\begin{widetext}
\begin{equation}
\label{eq:asymmetricpair}
pp\to\left\{\begin{array}{lclcl}
  \widetilde{R}^{+2/3}_2 \widetilde{R}^{-2/3}_2,  \widetilde{R}^{-2/3}_2 \widetilde{R}^{-2/3}_2 \\
  \widetilde{R}^{+2/3}_2 \widetilde{R}^{+2/3}_2,  \widetilde{R}^{+2/3}_2 \widetilde{R}^{-1/3}_2 \\
  \widetilde{R}^{+2/3}_2 \widetilde{R}^{+1/3}_2,  \widetilde{R}^{-2/3}_2 \widetilde{R}^{-1/3}_2 \\
  \widetilde{R}^{-2/3}_2 \widetilde{R}^{+1/3}_2 \\
\end{array}\right\} \rightarrow (j N) \, (j \mu) \rightarrow j(\mu j j)\ j\ \mu \equiv \mu^{\pm} \mu^{\pm} + N_{jet} \ge 4
\end{equation}    
\end{widetext}

\begin{center}
\textbf{Asymmetric mode: Single production}
\end{center}
\begin{widetext}
\begin{equation}
\label{eq:asymmetricsingle}
pp\to\left\{\begin{array}{lclcl}
\widetilde{R}^{+2/3}_2 \mu j,\ \widetilde{R}^{-2/3}_2 \mu j &\rightarrow& (j N)\, \mu j,\ (j N)\, \mu j &\rightarrow& j (\mu j j)\ j \mu \equiv \mu^{\pm} \mu^{\pm} + N_{jet} \ge 4 \\
\widetilde{R}^{+2/3}_2 j N,\ \widetilde{R}^{-2/3}_2 j N &\rightarrow& (j \mu) \ j N, \ (j \mu) \ j N  &\rightarrow& j \mu j (\mu j j)\ \equiv \mu^{\pm} \mu^{\pm} + N_{jet} \ge 4 \\
\end{array}\right\}.
\end{equation}    
\end{widetext}

\begin{figure*}

\includegraphics[width=10.00cm,height=7.00cm]{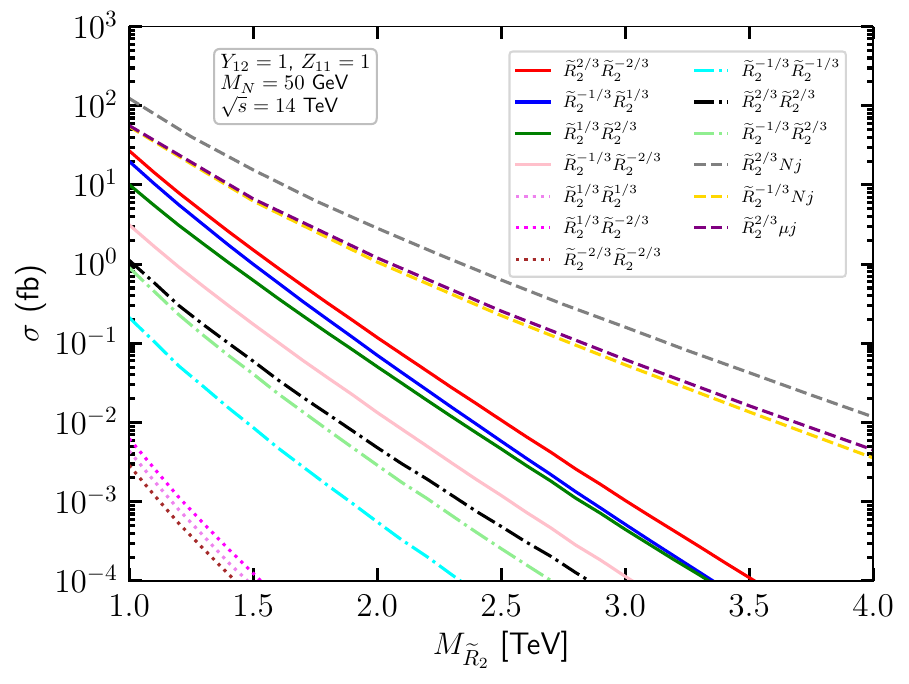}
\captionsetup{justification=raggedright,singlelinecheck=false}
\caption{Variation of the production cross section for different $\widetilde{R}_2$ pair and single production channels as a function of $M_{\widetilde{R}_2}$ at HL-LHC with C.O.M energy $\sqrt{s}=14$~TeV. The Yukawa couplings are fixed to $Z_{11}=Y_{12}=1$. The legend in the figure identifies the individual production modes.}
\label{fig:crossSection_single}
\end{figure*}

\noindent
In Fig.~\ref{fig:crossSection_single}, we present the production cross sections as functions of $M_{\widetilde{R}_2}$ for the channels discussed above at the HL-LHC, assuming the benchmark choice $Y_{12}=Z_{11}=1$. Throughout the figure, we show only the parton-level production cross sections of $\widetilde{R}_2$. The salient features of the plot are discussed below:
\begin{itemize}
    \item[--] 
    The pair production channels $\widetilde{R}^{2/3}_2 \widetilde{R}^{-2/3}_2$ and $\widetilde{R}^{-1/3}_2 \widetilde{R}^{1/3}_2$ are shown by the solid red and blue curves, respectively. Unlike ~\cite{Saha:2025npi}, where the SM mediated, new-physics mediated, and interference contributions were displayed separately, here we present the total cross section for each channel as a single contour. We observe that the cross section for the $\widetilde{R}^{2/3}_2 \widetilde{R}^{-2/3}_2$ channel is slightly larger than that for $\widetilde{R}^{-1/3}_2 \widetilde{R}^{1/3}_2$. The SM and new-physics contributions involving a $t$-channel $\mu$ ($\nu_{\mu}$) mediator for $\widetilde{R}^{2/3}_2$ ($\widetilde{R}^{-1/3}_2$) pair production, as well as the corresponding interference terms, are identical for these two modes. The observed difference arises from the pair production diagrams mediated by a $t$-channel $N$. In this case, the initial-state quark is an up quark for $\widetilde{R}^{2/3}_2$ production and a down quark for $\widetilde{R}^{-1/3}_2$ production. Since the parton distribution function (PDF) of the up quark inside the proton is significantly larger than that of the down quark, the contribution from this diagram enhances the cross section for $\widetilde{R}^{2/3}_2 \widetilde{R}^{-2/3}_2$ production.

   \item[--] 
   The pair production cross sections exhibit a well-defined hierarchy among the different charge combinations, with the largest contribution arising from $\widetilde{R}^{2/3}_2 \widetilde{R}^{2/3}_2$, followed by $\widetilde{R}^{-1/3}_2 \widetilde{R}^{2/3}_2$, $\widetilde{R}^{-1/3}_2 \widetilde{R}^{-1/3}_2$, $\widetilde{R}^{1/3}_2 \widetilde{R}^{-2/3}_2$, $\widetilde{R}^{1/3}_2 \widetilde{R}^{1/3}_2$, and finally $\widetilde{R}^{-2/3}_2 \widetilde{R}^{-2/3}_2$. This hierarchy can be understood from the structure of the initial-state partons. The $\widetilde{R}^{2/3}_2 \widetilde{R}^{2/3}_2$ production cross section is the most dominant because it arises from $t$-channel Majorana RHN exchange involving up quarks as the initial partons. Similarly, the mixed-charge channel $\widetilde{R}^{-1/3}_2 \widetilde{R}^{2/3}_2$ receives sizable contributions from both $s$-channel electroweak processes and $t$-channel RHN exchange, with at least one up-quark in the initial state, resulting in a relatively large cross section. For $\widetilde{R}^{-1/3}_2 \widetilde{R}^{-1/3}_2$, the production is dominated by down-quark–initiated processes, which are suppressed relative to up-quark–initiated channels due to the smaller down-quark PDF. For $\widetilde{R}^{-1/3}_2 \widetilde{R}^{-1/3}_2$, the production is dominated by down-quark–initiated processes, which are suppressed relative to up-quark–initiated channels due to the smaller down-quark PDF. The channels $\widetilde{R}^{1/3}_2 \widetilde{R}^{-2/3}_2$ and $\widetilde{R}^{1/3}_2 \widetilde{R}^{1/3}_2$ receive contributions from initial states involving a combination of quarks and anti-quarks or purely anti-quark–initiated processes, leading to further suppression. Finally, the $\widetilde{R}^{-2/3}_2 \widetilde{R}^{-2/3}_2$ channel is the most suppressed, as it predominantly involves anti-up quarks in the initial state, whose PDFs are significantly smaller.

    \item [--] In this section, we identify three primary single production channels contributing to the targeted signal topology: $\widetilde{R}^{2/3}_2 N j$, $\widetilde{R}^{-1/3}_2 N j$, and $\widetilde{R}^{2/3}_2 \mu j$. The corresponding Feynman diagrams, focusing on the dominant quark- and gluon-initiated subprocesses, are illustrated in Figs.~\ref{fig:R2single1}, \ref{fig:R2single3}, \ref{fig:R2single4}, and \ref{fig:R2single6}. Among these, the $\widetilde{R}^{2/3}_2 N j$ channel exhibits the largest cross section, exceeding that of $\widetilde{R}^{-1/3}_2 N j$ by approximately a factor of two. This disparity is driven by the proton's PDFs; specifically, the $\widetilde{R}^{2/3}_2 N j$ mode is initiated by up-quarks, whereas the $\widetilde{R}^{-1/3}_2 N j$ mode originates from down-quarks. The former is thus enhanced by the $u$-quark PDF at the LHC. We further note that the cross section for the $\widetilde{R}^{2/3}_2 \mu j$ channel follows a numerical trend comparable to that of $\widetilde{R}^{-1/3}_2 N j$.

\end{itemize}


\section{Search strategy for $\widetilde{R}_{2}$ at the HL-LHC} 
\label{sec:5}

\begin{figure*}[ht]
    \centering
    \captionsetup[subfigure]{labelformat=parens, labelfont=bf}

    \subfloat[]{
        \includegraphics[width=0.48\textwidth, height=0.35\textwidth]{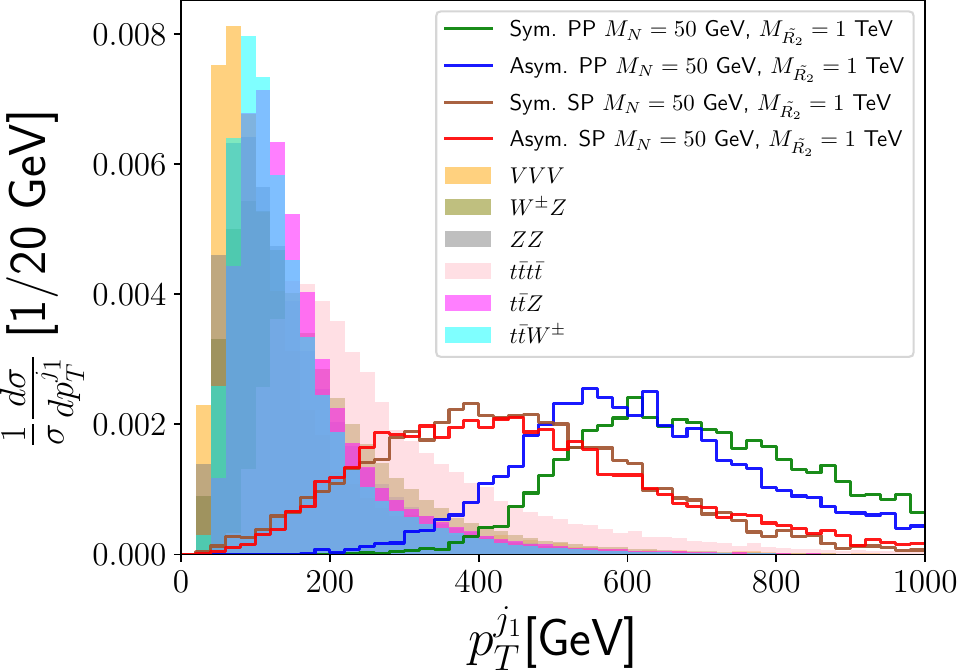}
       
    \label{fig:PTj1_SS}
    }
    \hfill
    \subfloat[]{
        \includegraphics[width=0.48\textwidth, height=0.35\textwidth]{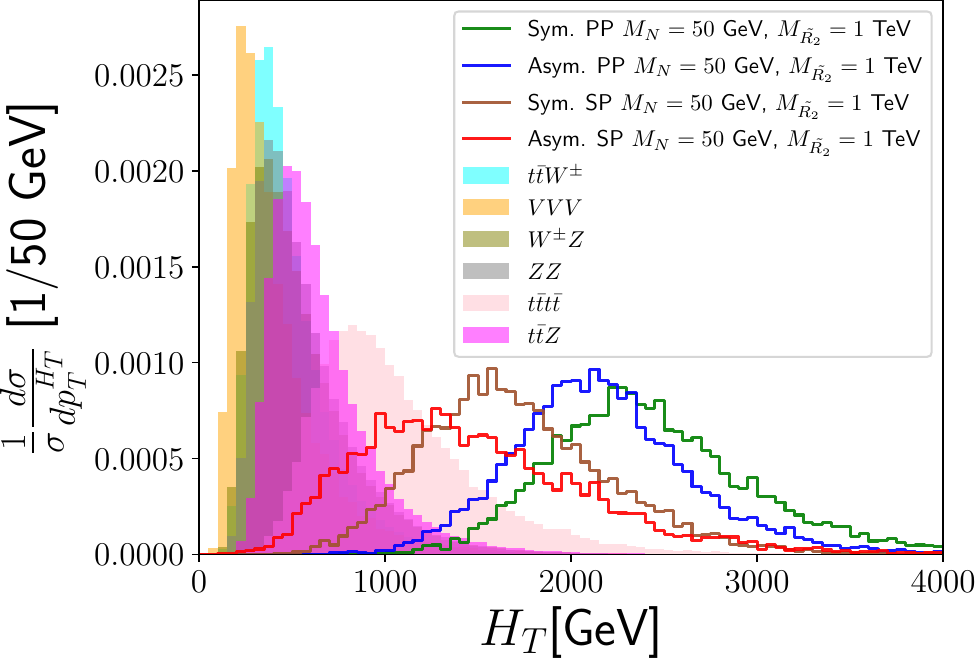}
       
    \label{fig:HT_SS}
    }
\captionsetup{justification=raggedright,singlelinecheck=false} 
    \caption{Distributions of (a) the transverse momentum of the leading jet ($p_{T}^{j_{1}}$) and (b) $H_T$ for benchmark values of $M_{\widetilde{R}_{2}^{2/3}} = 1$ TeV and $M_N = 50$ GeV in HL-LHC. }
    \label{fig:kinematicdistribution_direct}
\end{figure*}

\noindent
The simulation pipeline is similar to what we had used in our previous work~\cite{Saha:2025npi}. In this section we discuss the kinematic distribution of the signal and background processes and explain the background processes in detail in the following section. In Fig.~\ref{fig:kinematicdistribution_direct}, we present the kinematic distributions of the transverse momentum of the leading jet ($p_T^{j_1}$) and ($H_T$) for both signal and background processes. Here, $H_T$ is defined as the scalar sum of the transverse momenta of all visible particles,
\begin{equation}
H_T = \sum_{\text{jets}} p_T^{\text{jet}} \;+\; \sum_{\text{muons}} p_T^{\mu}.
\end{equation}
The choice of these kinematic observables is motivated by the clear separation they provide between signal and background event distributions. We present the distributions separately for the symmetric and asymmetric modes of LQ production, including both pair and single production, for the benchmark scenario with $M_{\widetilde{R}^{2/3}_2} = 1~\text{TeV}$ and $M_N = 50~\text{GeV}$. The kinematic distributions for $\widetilde{R}^{\mp 1/3}_2$ are largely similar to those of $\widetilde{R}^{\pm 2/3}_2$, except in the case of asymmetric production where they do not contribute. For this reason, we do not present them separately. \\

\begin{itemize}
    \item [--]

    The green and blue curves in Fig.~\ref{fig:PTj1_SS} correspond to the symmetric and asymmetric pair production modes, respectively. In both production modes, the leading jet originates from the decay of a $\widetilde{R}^{2/3}_2$. Consequently, its transverse momentum distribution exhibits a peak around $M_{\widetilde{R}^{2/3}_2}/2$, which for $M_{\widetilde{R}^{2/3}_2}=1$~TeV corresponds to approximately $500$~GeV. This feature is common to both the symmetric and asymmetric production modes.

    \item [--] The brown and red curves in Fig.~\ref{fig:PTj1_SS} represent the symmetric and asymmetric single production modes, respectively. In single production, only one heavy particle with mass $M_{\widetilde{R}_2} = 1$~TeV is produced, while the remaining energy in the event is shared with a relatively light recoil system. This redistribution of energy among lighter final-state particles leads to a softer $p_T$ spectrum for the leading jet compared to pair production, where the leading jet predominantly arises from the decay of a second, equally heavy $\widetilde{R}_2$. 
\end{itemize}

\begin{itemize} 
    \item [--] In figure~\ref{fig:HT_SS}, we present the $H_T$ distributions for the pair production processes, where the symmetric and asymmetric production modes are shown by the green and blue curves, respectively. In the symmetric production mode, both LQs are produced with similar transverse momentum. 
    Since the produced RHN  decays  entirely into visible final states, nearly the full energy of each LQ is transferred to jets and a muon. As can be seen from the figure, this results in a peak in $H_T$ distribution around $2 M_{\widetilde{R}_2} \simeq 2~\mathrm{TeV}$, with the maximum occurring around $2.3~\mathrm{TeV}$.  
    In contrast, for the asymmetric production mode, the $H_T$ distribution peaks at a slightly lower value, around $2.1~\mathrm{TeV}$. This reduction can be understood from the fact that only one of the produced LQs undergoes the decay chain involving the RHN. As a result, the visible final-state multiplicity is reduced, leading to a comparatively smaller overall scalar sum.

    \item[--] The brown and red curves in figure~\ref{fig:HT_SS} represent the $H_T$ distributions for the symmetric and asymmetric single production modes, respectively. In the symmetric single production case, the $H_T$ distribution peaks at a significantly lower value, around $1.6~\mathrm{TeV}$. This behaviour is expected, since only one heavy LQ is produced, and therefore the visible transverse momentum is primarily governed by a single $\sim 1~\mathrm{TeV}$ resonance together with its recoil momentum. The associated RHN  and the accompanying recoil jet typically carry comparatively softer transverse momenta, resulting in a substantially smaller overall scalar sum than in the pair production scenarios. In the asymmetric single production mode, the reduction in the number of visible final-state particles further suppresses the total transverse activity. Consequently, the $H_T$ distribution peaks at an even lower value, around $1.2~\mathrm{TeV}$. 
   
\end{itemize}

After analyzing the kinematic distributions shown in Fig.~\ref{fig:kinematicdistribution_direct}, we implement the following set of cuts, which provide effective discrimination between signal and background processes.

\begin{itemize}
\item The number of muons and jets: $N_{\text{muon}} = 2$ with same charge and $N_{\text{jet}} \geq 4$. We do not demand any specific isolation criterion for the muon. We apply base level cut on the muon transverse momentum, $p^{\mu}_T=20$ GeV.

\item 
$b$ veto: number of $b$-jets $N_b=0$ to suppress $t \bar{t}$ backgrounds.

\item Among the jets, $N_{\text{jets}}(p_T^j > 200~\text{GeV},~\Delta R_{\mu j} > 0.4) \geq 2$.
\item $H_T > 2000$ GeV.

\end{itemize}


\subsection*{Background processes}

\noindent
In this section, we enumerate the SM background processes that can mimic the signal topology of same sign dimuons accompanied by multiple jets. The relevant background categories are summarized below.
\begin{enumerate}
    \item $VVV$: 
    Triboson production constitutes one of the dominant SM backgrounds for our signal topology. The relevant triboson combinations include $Z_{\ell}Z_{\ell}Z_{h}$, $W_{\ell}W_{h}Z_{\ell}$, $W_{\ell}W_{\ell}W_{\ell}$, and $W_{\ell}Z_{\ell}Z_{h}$. Here, the subscripts $\ell$ and $h$ denote leptonic decays ($W \to \mu \nu_{\mu}$ or $Z \to \mu^+ \mu^-$) and hadronic decays ($V \to jj$) of the gauge bosons, respectively. These processes typically yield more than two muons in the final state and can therefore contribute to same sign dimuon signatures. The accompanying multiple jets arise either from hadronic gauge-boson decays or from additional QCD radiation during parton showering and hadronization.
    
    \item $VV+\text{jets}$: 
    In the diboson plus jets category, the dominant background contributions arise from the processes $W_{\ell}Z_{\ell}+\text{jets}$ and $Z_{\ell}Z_{\ell}+\text{jets}$. The leptonic decays of the gauge bosons produce multiple muons in the final state, while the required jet multiplicity comes from the additional jets.
    
    \item $t\bar{t}V$: 
    This background category includes two distinct classes of processes. The first consists of $t_{\ell}\bar{t}_{h}W^+_{\ell}$ and $t_{h}\bar{t}_{\ell}W^-_{\ell}$. In these channels, the leptonic decay of the top (antitop) quark produces a $b$ ($\bar b$) quark and a $W^+$ ($W^-$) boson, with the $W$ subsequently decaying leptonically to $\mu^+\nu_{\mu}$ ($\mu^-\bar{\nu_{\mu}}$). The additional $W^+$ ($W^-$) boson produced in association with the $t\bar{t}$ pair ensures the presence of same sign dimuons in the final state, while the hadronic decay of the remaining top quark provides additional jets. 
    The second class corresponds to the process $t_{h}\bar{t}_{\ell}Z_{\ell}$. In this case, the leptonic decays of the top quark and the $Z$ boson yield atleast three muons in the final state, which can mimic the signal topology upon appropriate event selection. The required jet activity arises from the hadronic decay of the remaining top quark and additional QCD radiation.
    
    \item $t\bar{t}t\bar{t}$: 
    We consider all possible decay configurations of the four top quarks that can contribute to the signal-like final state. Purely hadronic decay modes, such as $t_{h}\bar{t}_{h}t_{h}\bar{t}_{h}$, do not produce leptons and therefore do not constitute a relevant background. In contrast, decay modes such as $t_{h}\bar{t}_{\ell}t_{\ell}\bar{t}_{\ell}$ lead to multiple muons in the final state and can mimic the same sign dimuon signature. Similarly, configurations like $t_{\ell}\bar{t}_{h}t_{\ell}\bar{t}_{h}$ can also produce same sign dimuons, with the accompanying jets arising from the hadronic decays of the remaining top quarks.
\end{enumerate}
In addition to the dominant backgrounds discussed above, diboson ($VV$) and ditop ($t\bar{t}$) processes can, in principle, contribute to the same-sign dilepton signal topology through charge misidentification. For instance, diboson channels such as $W_h Z_{\ell}$ and $Z_h Z_{\ell}$ may yield opposite-sign dileptons at parton level; however, if the charge of one lepton is misidentified, these events can mimic a same-sign dilepton final state. A similar situation arises in ditop production, where configurations such as $t_{\ell}\bar{t}_{\ell}$ produce opposite-sign leptons, but charge misreconstruction of one lepton could lead to an apparent same-sign signature. {In our analysis, we have not included such effects.}  After applying realistic charge identification criteria and incorporating the corresponding misidentification probabilities, we find that the resulting contribution from these channels is negligible. Furthermore, the implementation of a $b$-jet veto substantially suppresses the ditop background, rendering its residual impact on the signal region insignificant.


\section{Analysis for HL-LHC} 
\label{sec:6}
\noindent
In this section, we describe the methodology used to calculate the significance of the signal topology. For each case, the statistical significance $\mathcal{Z}$ is calculated using the following expression outlined in~\cite{Cowan:2010js},

\begin{align}
\mathcal{Z} = \sqrt{2\left(N_S+N_B\right)\ln\left(\frac{N_S+N_B}{N_B}\right)-2N_S}\, ,
\label{eq:sig}
\end{align}
Here, $N_S$ and $N_B$ are the number of signal and background events, respectively.
The background events are computed as follows,
\begin{align}
\label{eq:Numevents_bkg}
N_B = (\sum_{i}\sigma_{B}^{i} \times \epsilon_{B}^{i}) \times \mathcal{L},    
\end{align}
\noindent
Here, $\sigma_{B}^{i}$ and $\epsilon_{B}^{i}$ denote the cross section and cut efficiency of the $i^{th}$ background process, respectively. $\mathcal{L}$ is the luminosity of the HL-LHC. The number of signal events $N_S$ is parameterized in the following manner. $N_S$ comprises of events from pair (symmetric and asymmetric) and single production (symmetric and asymmetric) modes. The number of signal events is given as, 

\begin{align}
\label{eq:Numevents_signal}
N_S = N^{\rm pair}_{\rm sym} + N^{\rm single}_{\rm sym} + N^{\rm pair}_{\rm asym} + N^{\rm single}_{\rm asym}, 
\end{align}
Here, $N^{\rm pair}_{\rm sym}$ ($N^{\rm pair}_{\rm asym}$) and $N^{\rm single}_{\rm sym}$ ($N^{\rm single}_{\rm asym}$) correspond to signal events from the symmetric (asymmetric) pair and single production modes, respectively. We write $N^{\rm pair}_{\rm sym}$, $N^{\rm pair}_{\rm asym}$, $N^{\rm single}_{\rm sym}$, and $N^{\rm single}_{\rm asym}$ as follows,

\begin{align}
\label{eq:Numevents_pair_sym}
N^{\rm pair}_{\rm sym} = \sigma^{\rm pair}_{\rm sym} \times \epsilon_{\rm sym}^{\rm pair} \times \beta^2(\widetilde{R}_2\rightarrow N j)\times \beta^2(N\rightarrow \mu j j) \times \mathcal{L}
\end{align}

\begin{align}
\label{eq:Numevents_pair_asym}
N^{\rm pair}_{\rm asym} = 2\times \sigma^{\rm pair}_{\rm asym} \times \epsilon_{\rm asym}^{\rm pair} \times \beta(\widetilde{R}_2\rightarrow \mu j) \times \beta(\widetilde{R}_2\rightarrow N j)\newline\\
\times \beta(N\rightarrow \mu j j) \times \mathcal{L}
\end{align}

\begin{align}
\label{eq:Numevents_single_sym}
N^{\rm single}_{\rm sym} = \sigma^{\rm single}_{\rm sym} (pp\rightarrow\widetilde{R}_2 Nj) \times \epsilon_{\rm sym}^{\rm single} \times \beta(\widetilde{R}_2\rightarrow N j)\newline\\
\times \beta^2(N\rightarrow \mu j j) \times \mathcal{L}
\end{align}

\begin{align}
\label{eq:Numevents_single_asym}
N^{\rm single}_{\rm asym} = \sigma^{\rm single}_{\rm asym} (pp\rightarrow\widetilde{R}_2 Nj/\widetilde{R}_2\mu j) \times \epsilon_{\rm asym}^{\rm single}\newline\\
\times \beta(\widetilde{R}_2\rightarrow \mu j/N j)\times \beta(N\rightarrow \mu j j) \times \mathcal{L}
\end{align}
\noindent
Here, $\beta$ corresponds to the BR. $\epsilon^{\rm pair}_{\rm sym}$ ($\epsilon^{\rm pair}_{\rm asym}$) and $\epsilon^{\rm single}_{\rm sym}$ ($\epsilon^{\rm single}_{\rm asym}$) are the efficiencies of the symmetric (asymmetric) pair and single production modes, respectively.
 
In Eqs.~\ref{eq:symmetricpair} and \ref{eq:asymmetricpair}, we have listed all the possible pair production channels of $\widetilde{R}_2$, whose decay modes can lead to the desired final state.  
The cross section parameterization of the symmetric pair production modes is as follows,
\begin{align}
\label{eq:symPP}
\sigma^{\rm pair}_{\rm sym} = \sigma(\widetilde{R}^{2/3}_2\widetilde{R}^{-2/3}_2) + \sigma(\widetilde{R}^{-2/3}_2\widetilde{R}^{-2/3}_2) + \sigma(\widetilde{R}^{2/3}_2\widetilde{R}^{2/3}_2) \newline \\
+ \sigma(\widetilde{R}^{1/3}_2\widetilde{R}^{-1/3}_2) + \sigma(\widetilde{R}^{1/3}_2\widetilde{R}^{1/3}_2)+ \sigma(\widetilde{R}^{-1/3}_2\widetilde{R}^{-1/3}_2) \newline \\
+ \sigma(\widetilde{R}^{2/3}_2\widetilde{R}^{-1/3}_2) + \sigma(\widetilde{R}^{2/3}_2\widetilde{R}^{1/3}_2)
+ \sigma(\widetilde{R}^{-2/3}_2\widetilde{R}^{-1/3}_2) \newline \\
 + \sigma(\widetilde{R}^{-2/3}_2\widetilde{R}^{1/3}_2) 
\end{align}
\noindent
and the parameterization of the asymmetric pair production modes is as follows, 

\begin{align}
\label{eq:asymPP}
\sigma^{\rm pair}_{\rm asym} = \sigma(\widetilde{R}^{2/3}_2\widetilde{R}^{-2/3}_2) + \sigma(\widetilde{R}^{-2/3}_2\widetilde{R}^{-2/3}_2) + \sigma(\widetilde{R}^{2/3}_2\widetilde{R}^{2/3}_2) \newline \\
+ \sigma(\widetilde{R}^{2/3}_2\widetilde{R}^{-1/3}_2) + \sigma(\widetilde{R}^{2/3}_2\widetilde{R}^{1/3}_2)
 \newline \\ 
+ \sigma(\widetilde{R}^{-2/3}_2\widetilde{R}^{1/3}_2)
\end{align}

We note that in the asymmetric production mode, we do not consider $\widetilde{R}^{\pm 1/3}_2 \widetilde{R}^{\pm 1/3}_2 $ and $\widetilde{R}^{ 1/3}_2 \widetilde{R}^{- 1/3}_2 $ modes, as they will lead to final states involving missing energy from the decay $\widetilde{R}^{-1/3}_2 \to d \nu$, and not same-sign di-muon.
The individual cross section are parameterized as follows,
\begin{align}
\label{eq:PP23}
\sigma(\widetilde{R}^{2/3}_2\widetilde{R}^{-2/3}_2) = 
\sigma^{\rm EW}_{\widetilde{R}^{2/3}_2} 
+ \sigma^{\rm QCD}_{\widetilde{R}^{2/3}_2} 
+ Y_{12}^{2}\Big(\sigma^{\rm Int,\,QCD}_{\mu,\,\widetilde{R}^{2/3}_2}
+ \sigma^{\rm Int,\,EW}_{\mu,\,\widetilde{R}^{2/3}_2}
\Big) \newline \\
+ Y_{12}^{4}\, \sigma^{\mu}_{\widetilde{R}^{2/3}_2}
+ Z_{11}^{2}\Big(\sigma^{\rm Int,\,QCD}_{N,\,\widetilde{R}^{2/3}_2}
  + \sigma^{\rm Int,\,EW}_{N,\,\widetilde{R}^{2/3}_2}
\Big) \newline \\
+ Z_{11}^{4} \sigma^{N}_{\widetilde{R}^{2/3}_2}
+ Z_{11}^{2}Y_{12}^{2}\sigma^{\rm Int}_{\mu N,\,\widetilde{R}^{2/3}_2} &
\end{align}


\begin{align}
\label{eq:PP13}
\sigma(\widetilde{R}^{1/3}_2\widetilde{R}^{-1/3}_2) =
\sigma^{\rm EW}_{\widetilde{R}^{-1/3}_2} 
+ \sigma^{\rm QCD}_{\widetilde{R}^{-1/3}_2} 
+ Y_{12}^{2}\Big(
    \sigma^{\rm Int,\,QCD}_{\nu,\,\widetilde{R}^{-1/3}_2}
  + \sigma^{\rm Int,\,EW}_{\nu,\,\widetilde{R}^{-1/3}_2}
\Big) \newline \\
+ Y_{12}^{4}\, \sigma^{\nu}_{\widetilde{R}^{-1/3}_2}
+ Z_{11}^{2}\Big(
    \sigma^{\rm Int,\,QCD}_{N,\,\widetilde{R}^{-1/3}_2}
  + \sigma^{\rm Int,\,EW}_{N,\,\widetilde{R}^{-1/3}_2}
\Big)\newline \\ + Z_{11}^{4} \sigma^{N}_{\widetilde{R}^{-1/3}_2}
+ Z_{11}^{2}Y_{12}^{2} \sigma^{\rm Int}_{\nu N,\,\widetilde{R}^{-1/3}_2} .
\end{align}

\begin{align}
\label{eq:PP2313}
\sigma(\widetilde{R}^{\pm 2/3}_2\widetilde{R}^{\mp 1/3}_2) = 
\sigma^{\rm EW}_{\widetilde{R}^{\pm 2/3}_2\widetilde{R}^{\pm 1/3}_2} 
+ Z_{11}^{2}\, \sigma^{\rm Int,\,EW}_{N,\widetilde{R}^{\pm 2/3}_2\widetilde{R}^{\mp 1/3}_2} \newline \\
+ Z_{11}^{4}\, \sigma^{N}_{\widetilde{R}^{\pm 2/3}_2\widetilde{R}^{\mp 1/3}_2} 
\end{align}


\begin{align}
\label{eq:PP223213all1}
\sigma(\widetilde{R}^{\pm 2/3}_2\widetilde{R}^{\pm 2/3}_2) &= Z_{11}^{4}\, \sigma^{N}_{\widetilde{R}^{\pm 2/3}_2\widetilde{R}^{\pm 2/3}_2} \\[4pt]
\label{eq:PP223213all2}
\sigma(\widetilde{R}^{\pm 1/3}_2\widetilde{R}^{\pm 1/3}_2) &= Z_{11}^{4}\, \sigma^{N}_{\widetilde{R}^{\pm 1/3}_2\widetilde{R}^{\pm 1/3}_2} \\[4pt]
\label{eq:PP223213all3}
\sigma(\widetilde{R}^{\pm 1/3}_2\widetilde{R}^{\pm 2/3}_2) &= Z_{11}^{4}\, \sigma^{N}_{\widetilde{R}^{\pm 1/3}_2\widetilde{R}^{\pm 2/3}_2} 
\end{align}

\begin{figure*}[ht]
\centering
\subfloat[]{\includegraphics[width=0.40\textwidth]{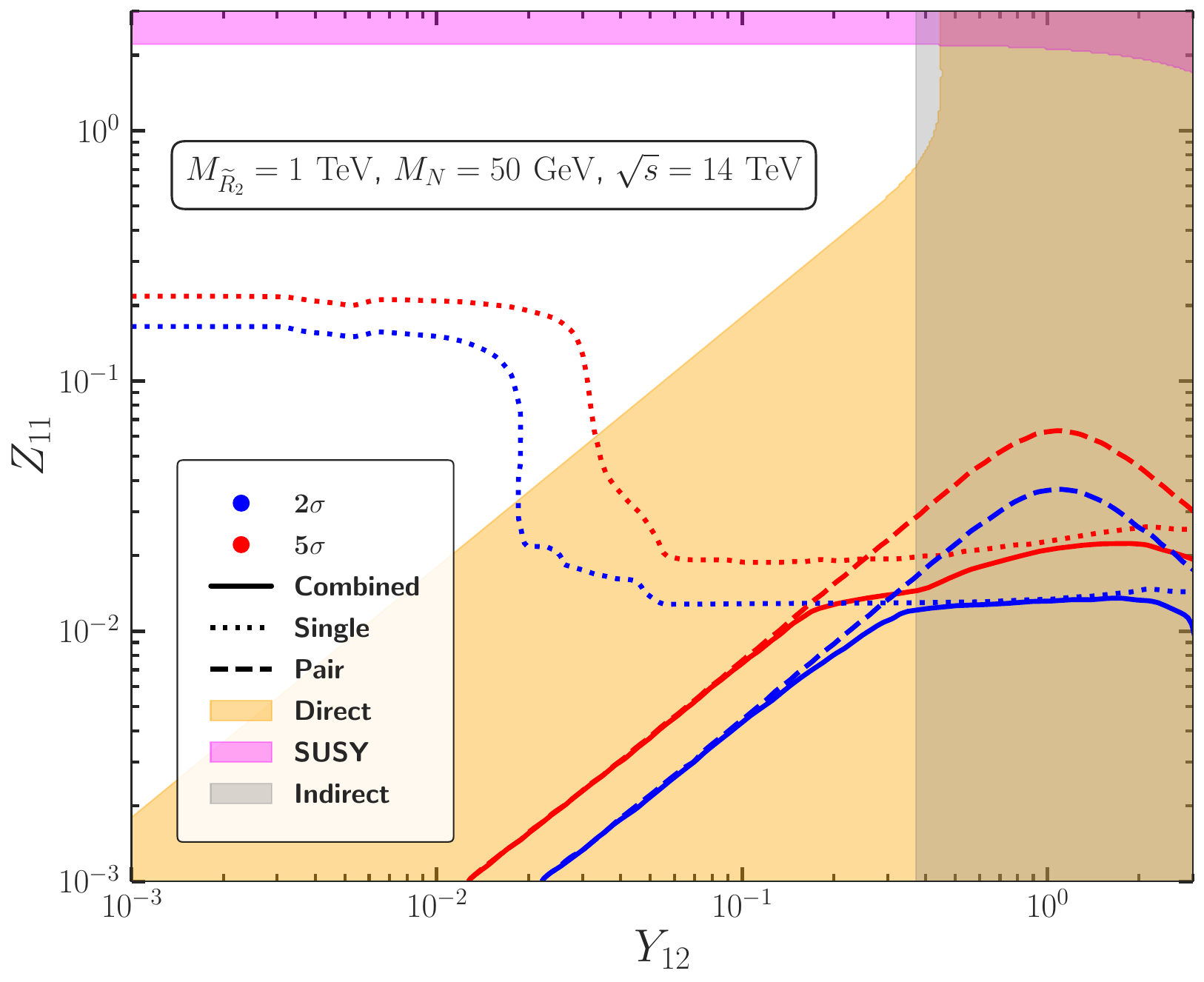}
\label{fig:combined_significance_1_TeV_MLQ_MN_50}}
\subfloat[]{\includegraphics[width=0.40\textwidth]{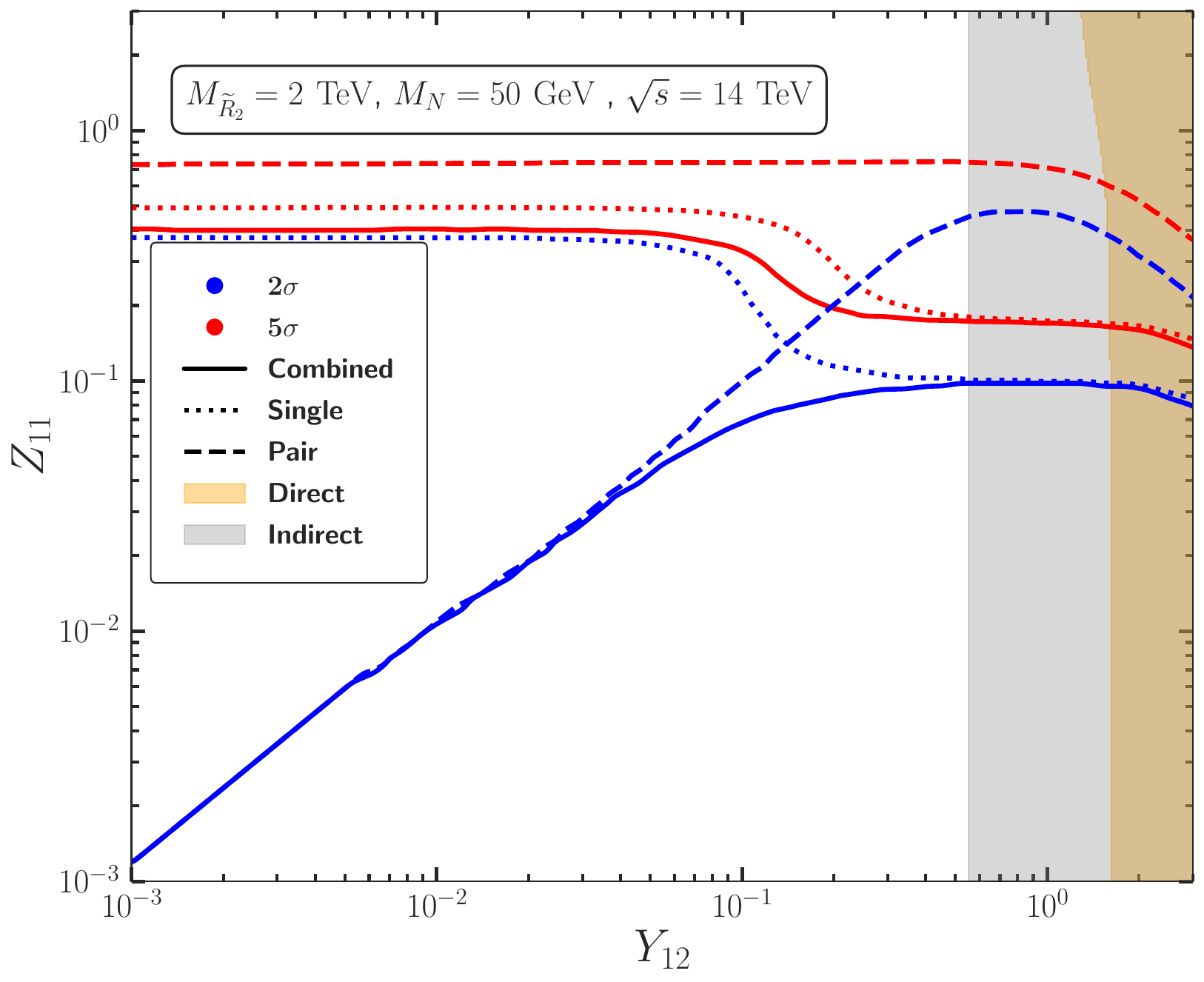}
\label{fig:combined_significance_2_TeV_MLQ_MN_50}}\\
\subfloat[]{\includegraphics[width=0.40\textwidth]{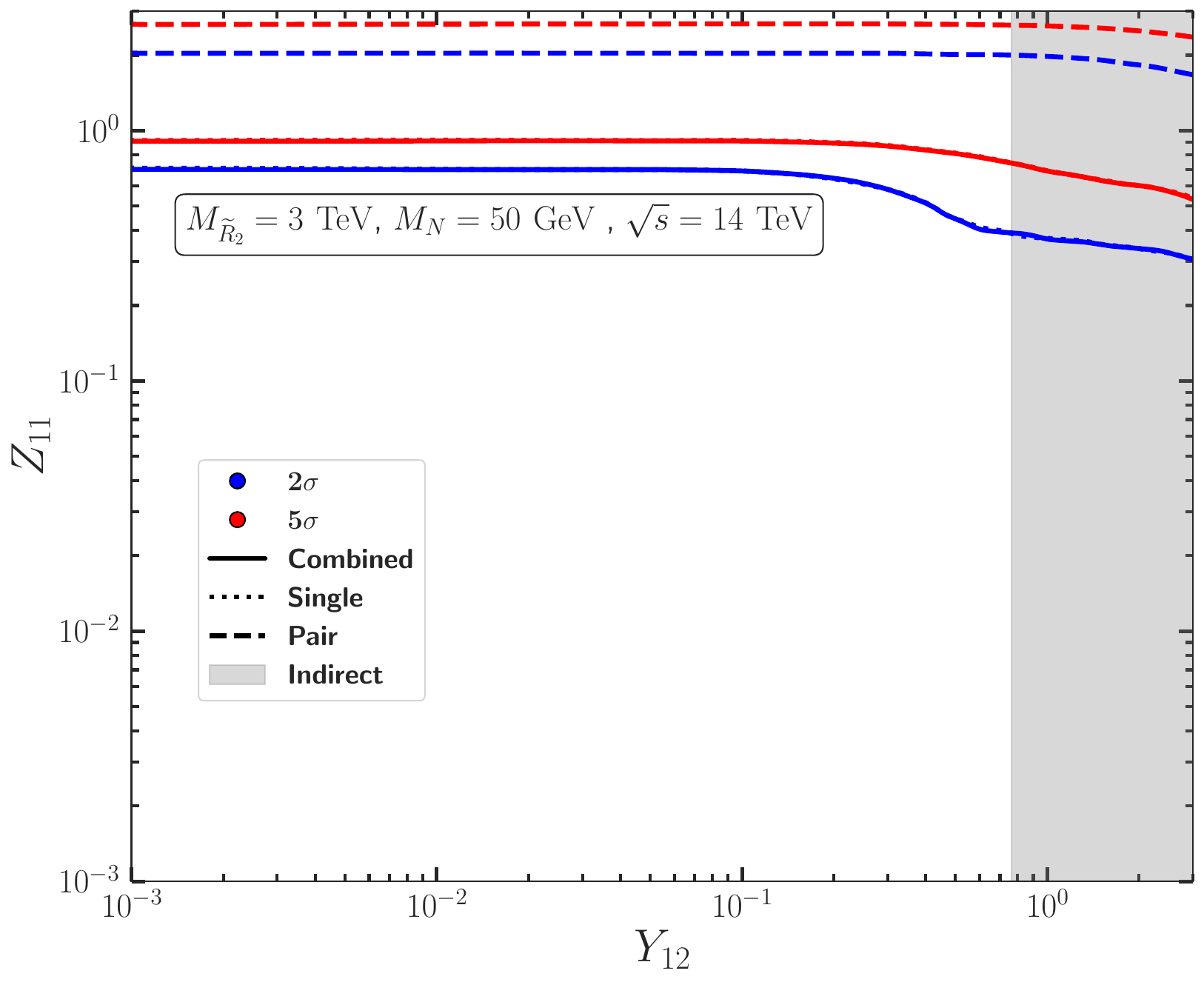}
\label{fig:combined_significance_3_TeV_MLQ_MN_50}}
\subfloat[]{\includegraphics[width=0.40\textwidth]{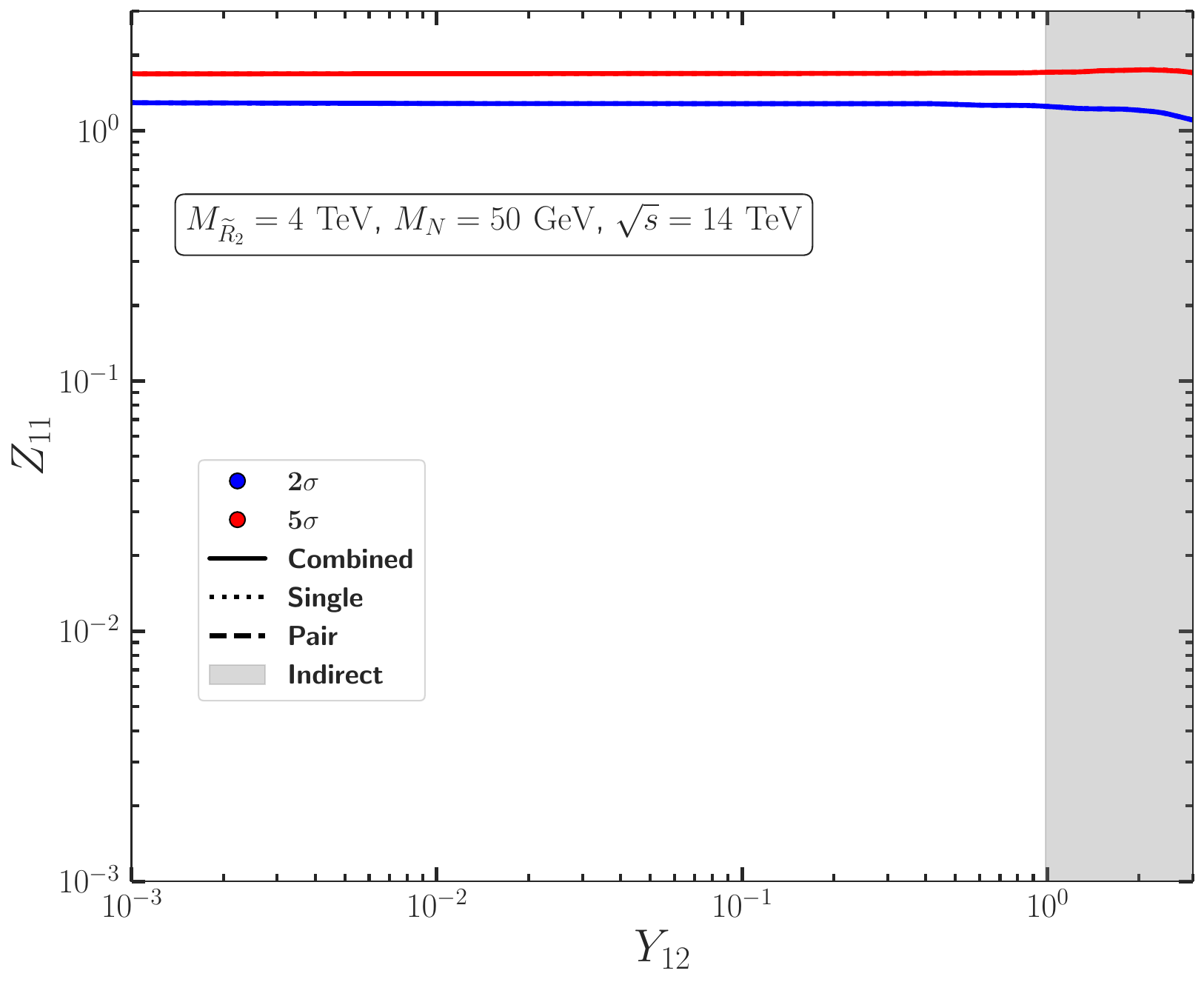}
\label{fig:combined_significance_4_TeV_MLQ_MN_50}}\\
\captionsetup{justification=raggedright,singlelinecheck=false} 
\caption{Here, we show the contour plots for $2\sigma$ and $5\sigma$ statistical significance  in the $Y_{12}-Z_{11}$ plane. We fix $M_N=50$ GeV. (a); The  mass of sLQ has been set to $\widetilde{R_2}=1.0$ TeV. The yellow region is excluded by ATLAS direct search~\cite{ATLAS:2020dsk}. The pink region is excluded by ATLAS SUSY gluino direct search~\cite{ATLAS:2023afl}. The gray region is excluded by the CMS indirect search~\cite{CMS:2024bej}. (b) Here we consider, mass of $\widetilde{R_2}=2.0$ TeV. (c) Here we consider, mass of $\widetilde{R_2}=3.0$ TeV. (d) Here we consider, mass of $\widetilde{R_2}=4.0$ TeV.}
\label{Fig:Ydl_MR2_final}
\end{figure*}
In the above equations, the various notations used for the production cross sections $\sigma$ are defined as follows:
\begin{itemize}
    \item[--] 
    $\sigma^{\rm EW}_{\widetilde{R}^{2/3}_2}$ ($\sigma^{\rm EW}_{\widetilde{R}^{-1/3}_2}$) and $\sigma^{\rm QCD}_{\widetilde{R}^{2/3}_2}$ ($\sigma^{\rm QCD}_{\widetilde{R}^{-1/3}_2}$) denote the pair production of $\widetilde{R}^{2/3}_2$ ($\widetilde{R}^{-1/3}_2$) mediated by an $s$-channel $Z/\gamma$ boson and a gluon, respectively.

    \item[--] 
    $\sigma^{\mu}_{\widetilde{R}^{2/3}_2}$ and $\sigma^{N}_{\widetilde{R}^{-1/3}_2}$ refer to the pair production of $\widetilde{R}^{2/3}_2$ and $\widetilde{R}^{-1/3}_2$ mediated via a $t$-channel exchange of a $\mu$ and a $N$, respectively. $\sigma^{N}_{\widetilde{R}^{2/3}_2}$ ($\sigma^{\nu}_{\widetilde{R}^{-1/3}_2}$) denotes the pair production of $\widetilde{R}^{2/3}_2$ ($\widetilde{R}^{-1/3}_2$) mediated via a $t$-channel exchange of a $N$ ($\nu$). 

    \item[--] $\sigma^{\rm Int,\,QCD}_{\mu,\,\widetilde{R}^{2/3}_2}$ ($\sigma^{\rm Int,\,EW}_{\mu,\,\widetilde{R}^{2/3}_2}$) denotes the interference contribution between the $t$-channel $\mu$-mediated pair production of $\widetilde{R}^{2/3}_2$ and the $s$-channel SM QCD (EW)-mediated pair production process. An analogous interpretation applies to $\sigma^{\rm Int,\,QCD}_{N,\,\widetilde{R}^{-1/3}_2}$ and $\sigma^{\rm Int,\,EW}_{N,\,\widetilde{R}^{-1/3}_2}$ ($\sigma^{\rm Int,\,QCD}_{\nu,\,\widetilde{R}^{-1/3}_2}, \sigma^{\rm Int,\,EW}_{\nu,\,\widetilde{R}^{-1/3}_2}$), with the only difference being that the $t$-channel mediator in this case is $N$ ($\nu$).

    \item[--] 
    In Eq.~\ref{eq:PP2313}, $\sigma^{\rm EW}_{\widetilde{R}^{\pm 2/3}_2 \widetilde{R}^{\mp 1/3}_2}$ denotes the pair production of $\widetilde{R}^{\pm 2/3}_2 \widetilde{R}^{\mp 1/3}_2$ mediated by an $s$-channel $W$ boson. The term $\sigma^{N}_{\widetilde{R}^{\pm 2/3}_2 \widetilde{R}^{\mp 1/3}_2}$ corresponds to pair production via a $t$-channel exchange of $N$. The quantity $\sigma^{\rm Int,\,EW}_{N,\,\widetilde{R}^{\pm 2/3}_2 \widetilde{R}^{\mp 1/3}_2}$ represents the interference contribution between these two processes. 

    \item[--] 
    The cross section contributions appearing in Eqs.~\ref{eq:PP223213all1}, \ref{eq:PP223213all2}, and \ref{eq:PP223213all3} arise exclusively from pair production processes mediated by a $t$-channel exchange of $N$.
\end{itemize}


\section{Results and discussions} 
\label{sec:7}
\noindent
In Fig.~\ref{Fig:Ydl_MR2_final}, we present the projected sensitivity contours in the $Y_{12}$–$Z_{11}$ plane for representative sLQ masses $M_{\widetilde{R}_2}=1.0,\ 2.0,\ 3.0,$ and $4.0~\text{TeV}$, assuming $M_N=50~\text{GeV}$ at the HL-LHC. The dashed, dotted, and solid contours correspond to pair production, single production, and their combination, respectively, while the blue (red) curves denote the $2\sigma$ sensitivity ($5\sigma$ discovery reach). Existing constraints from ATLAS direct LQ searches~\cite{ATLAS:2020dsk}, ATLAS gluino-pair SUSY searches~\cite{ATLAS:2023afl}, and indirect CMS limits~\cite{CMS:2024bej} are shown as shaded regions. The recasting procedure follows Ref.~\cite{Saha:2025npi}.

For $M_{\widetilde{R}_2}=1.0~\text{TeV}$ (Fig.~\ref{fig:combined_significance_1_TeV_MLQ_MN_50}), the sensitivity is dominated by pair production. The corresponding contours exhibit an approximately linear and monotonic dependence on $Y_{12}$ and $Z_{11}$ up to $Y_{12}\lesssim 1$, reflecting the QCD-dominated nature of the production cross section, which is largely insensitive to the Yukawa couplings. In this regime, the signal yield is primarily controlled by the branching ratios, which remain nearly constant, leading to the observed scaling. For $Y_{12}\gtrsim 1$, Yukawa-induced contributions proportional to $Y_{12}^4$ become relevant, resulting in a turnover of the contours, where smaller values of $Z_{11}$ suffice to maintain a fixed significance.

In contrast, single production exhibits a distinct parametric dependence. At small $Y_{12}\lesssim 10^{-2}$, the sensitivity is largely controlled by $Z_{11}$, as the dominant gluon–quark initiated processes are insensitive to $Y_{12}$. As $Y_{12}$ increases, the contours develop a characteristic dip driven by the enhancement of $\mathrm{BR}(\widetilde{R}_2\to \mu j)$, which reduces the required $Z_{11}$ for a fixed signal yield. At larger couplings, the contours flatten, indicating a weak residual dependence on $Y_{12}$. The combined sensitivity closely follows pair production at small couplings and transitions smoothly to the single production behavior at larger couplings, demonstrating the complementarity of the two production modes.

For $M_{\widetilde{R}_2}=2.0~\text{TeV}$ (Fig.~\ref{fig:combined_significance_2_TeV_MLQ_MN_50}), the impact of phase-space suppression becomes evident. While the $2\sigma$ contours retain a monotonic dependence, the $5\sigma$ reach from pair production requires $Z_{11}\sim \mathcal{O}(1)$ and becomes largely insensitive to $Y_{12}$ for $Y_{12}\lesssim 1$. In this regime, contributions from $Z_{11}$-dependent diagrams, enabled by the Majorana nature of the RHN, partially compensate for the reduced QCD cross section. For $Y_{12}\gtrsim 1$, Yukawa-induced contributions again lead to a downward turn in the contours. The single production contours remain qualitatively similar to the $1.0~\text{TeV}$ case but with a smoother dependence on $Y_{12}$. Notably, the combined $5\sigma$ sensitivity is now largely driven by single production, signaling the onset of the transition in the dominant production mechanism. The SUSY-inspired constraints no longer significantly impact the parameter space.

For heavier masses, $M_{\widetilde{R}_2}=3.0$ and $4.0~\text{TeV}$ (Figs.~\ref{fig:combined_significance_3_TeV_MLQ_MN_50} and \ref{fig:combined_significance_4_TeV_MLQ_MN_50}), the phenomenology changes qualitatively. The sensitivity is almost entirely driven by single production, with the combined contours overlapping with those from single production. Pair production contributes only at very large $Z_{11}$ and remains largely insensitive to $Y_{12}$. In this regime, the HL-LHC sensitivity extends well beyond current direct limits, with only indirect constraints remaining relevant.

Overall, our results demonstrate a clear transition from QCD-dominated pair production at low masses to Yukawa-driven single production at higher masses. This interplay significantly enhances the HL-LHC reach and highlights the importance of combining all the relevant production modes. In particular, the same-sign dimuon channel provides a powerful and complementary probe of heavy sLQs coupled to Majorana RHNs, enabling access to regions of parameter space that remain unconstrained by existing searches.
\\
 Note that, among the sLQs that couple to RHNs—$\widetilde{R}_2(\mathbf{3},\mathbf{2},1/6)$, $S_1(\mathbf{\bar{3}},\mathbf{1},1/3)$, and $\bar{S}_1(\mathbf{\bar{3}},\mathbf{1},-2/3)$~\cite{Dorsner:2016wpm}—we have focused on $\widetilde{R}_2$. While $S_1$ lacks the doublet's $\widetilde{R}_2^{-1/3}$ state, it reproduces similar contributions analogous to the $\widetilde{R}_2^{+2/3}$ component. Conversely, $\bar{S}_1$ lacks $t$-channel $\mu^{\pm}$ exchange in sLQ production as well as contributions to RHN decays such as $N \rightarrow \mu j j$. Nevertheless, their overall experimental signatures remain qualitatively similar. To capture all relevant production and decay channels within a minimal set of new degrees of freedom, we adopt $\widetilde{R}_2(\mathbf{3}, \mathbf{2}, 1/6)$ as our representative RHN-coupled sLQs.


\section{Conclusions} 
\label{sec:8}
\noindent
We have performed a comprehensive study of the sLQ $\widetilde{R}_2$ in scenarios where it couples to a lighter RHN, focusing on collider signatures that evade conventional LQ searches. In the regime $M_{\widetilde{R}_2} > M_N$, the decay $\widetilde{R}_2 \to N j$ can dominate, leading to a same-sign dimuon and multijet final state. This channel is particularly clean, benefits from low Standard Model backgrounds, and directly probes the lepton-number-violating nature of a Majorana RHN.

Our analysis reveals a clear and robust interplay between pair and single production mechanisms. At low masses, $M_{\widetilde{R}_2} \sim 1\text{--}2~\text{TeV}$, the sensitivity is driven by QCD-dominated pair production and exhibits a weak dependence on the Yukawa couplings. As the mass increases, phase-space suppression reduces the impact of pair production, and Yukawa-driven single production becomes increasingly important, eventually dominating the sensitivity in the multi-TeV regime. This transition, observed directly in the sensitivity contours, highlights the necessity of combining production modes to fully exploit the HL-LHC reach.

A comparison with existing constraints shows that the same-sign dimuon channel probes substantial regions of the $Y_{12}$–$Z_{11}$ parameter space that remain unconstrained by current direct searches, SUSY-searches, and indirect limits. In particular, for heavy sLQs where pair production is suppressed, single production maintains significant sensitivity, extending the reach of the HL-LHC well beyond present bounds.

Overall, our results demonstrate that $\widetilde{R}_2$ with sizable couplings to RHNs remain a viable and testable scenario at the HL-LHC. The search strategy developed here provides a concrete and complementary probe of both sLQ dynamics and the Majorana nature of RHNs. Extensions to other flavor structures, RHN mass regimes, or complementary final states offer promising directions to further enhance the discovery potential at the HL-LHC and future collider experiments.

Note that, the same-sign dimuon plus multijet signature is not unique to the sLQ scenario considered in this work; it also features prominently across a variety of BSM extensions, for example $U(1)$ gauge theories with heavy $Z^\prime$ bosons~\cite{Mohapatra:1980qe,Wetterich:1981bx,Padhan:2022fak} and Type-II Seesaw frameworks with doubly charged Higgs scalars~\cite{Konetschny:1977bn,Mohapatra:1980yp,Maharathy:2023dtp,Englert:2026eou}. While these competing scenarios yield similar visible final states, they can be systematically disentangled through kinematic and topological features. Specifically, resonance reconstructions—such as the invariant mass distributions of the $\mu jj$, $jj$, and $\mu^\pm\mu^\pm$ systems—combined with the transverse momentum ($p_{\mathrm{T}}$), pseudorapidity ($\eta$), and angular separation ($\Delta R$) of the leading jets, provide a clear discriminant to isolate our leptoquark signal from alternative BSM scenarios.



\section*{ACKNOWLEDGMENTS}
\noindent
The authors acknowledge the use of SAMKHYA: High-Performance Computing Facility provided by the Institute of Physics (IOP), Bhubaneswar and the two workstations provided by the Institute of
Physics, Bhubaneswar from the DAE APEX project for numerical computations.

\bibliography{bibitem}
\bibliographystyle{utphys}

\end{document}